\documentclass[12pt]{iopart}

\usepackage{graphicx}
\usepackage{setstack}
\usepackage{iopams}
\usepackage{color}

\begin{document}

\title[Active transport improves the precision of linear long distance molecular signalling]{Active transport improves the precision of linear long distance molecular signalling}

\author{Alja\v{z} Godec$^{\dagger,\ddagger}$ and Ralf Metzler$^{\dagger}$}
\address{$\dagger$ Institute of Physics \& Astronomy, University of Potsdam, 14776
Potsdam-Golm, Germany\\
$\ddagger$ National Institute of Chemistry, 1000 Ljubljana, Slovenia}
%$\sharp$ Department of Physics, Tampere University of Technology, FI-33101
%Tampere, Finland}

\begin{abstract}
Molecular signalling in living cells occurs at low copy numbers and is thereby
inherently limited by the noise imposed by thermal diffusion. The precision at
which biochemical receptors can count signalling molecules is intimately related
to the noise correlation time. In addition to passive thermal diffusion, messenger
RNA and vesicle-engulfed signalling molecules can transiently bind to molecular
motors and are actively transported across biological cells. Active transport is
most beneficial when trafficking occurs over large distances, for instance up to
the order of 1 metre in neurons. Here we explain how intermittent active transport
allows for faster equilibration upon a change in concentration triggered by
biochemical stimuli. Moreover, we show how intermittent active excursions induce
qualitative changes in the noise in effectively one-dimensional systems such as
dendrites. Thereby they allow for significantly improved signalling precision in
the sense of a smaller relative deviation in the concentration read-out by the
receptor. On the basis of linear response theory we derive the exact mean field
precision limit for counting actively transported molecules. We explain how
intermittent active excursions disrupt the recurrence in the molecular motion,
thereby facilitating improved signalling accuracy. Our results provide a deeper
understanding of how recurrence affects molecular signalling precision in
biological cells and novel diagnostic devices.
\end{abstract}

\section{Introduction}

In his seminal work on the reaction-rate theory in the diffusion-controlled limit
Smoluchowski established a quantitative connection between thermal fluctuations
in the form of molecular diffusion and a macroscopically observable time evolution
of the concentration of reactants and products \cite{Smol}. Some 60 years later
Berg and Purcell \cite{BergPurc} showed that thermal diffusion also limits the
accuracy of biochemical receptors and hence sets physical bounds to the precision
of cellular signalling. Namely, cellular signalling typically involves low copy
numbers of messenger molecules and is thereby inevitably subjected to appreciable
fluctuations in the count of molecular binding events at biochemical receptors
\cite{BergPurc,Bialek,Tkacik,Endres,GodecActive,Goodhill}. In a similar way
counting noise limits the precision and sensitivity of modern microscopic
diagnostic devices \cite{Diagnostics}. State-of-the-art single particle tracking
techniques indeed highlight the inherent stochasticity of such molecular
signalling events \cite{Stoch,Xie,Elf1,Elf2}. However, despite the significant
sample-to-sample fluctuations cellular signalling operates at remarkable precision
\cite{BialekBK,Alberts}. Inside living cells some signalling molecules, typically
entrapped in vesicles, do not only move by thermal diffusion alone but may also be
actively transported along cellular filaments by molecular motors \cite{Motors1,
Motors2} causing intermittent ballistic excursions \cite{Interm1}. Free molecules,
such as messenger RNA, may as well attach to motors \cite{mrna}, or proteins may
move in a directed fashion due to cytoplasmic drag \cite{haim}. Enhanced
spreading may finally be facilitated by cytoplasmic streaming \cite{streaming,
christine}. A practical way to incorporate active motion in the stochastic
dynamics of signalling molecules is the model of random intermittent
search \cite{Gleb,RMP} which was recently used to analyse reaction kinetics in
active media \cite{Olivier} and the speed and precision of receptor signalling
in 3-dimensional media \cite{GodecActive}.

\begin{figure}
\begin{center}
\includegraphics[width=12.cm]{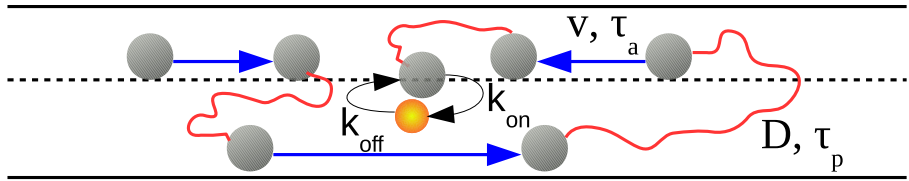}
\caption{Model system: signalling particles (grey) perform
passive thermal diffusion (red phases) interrupted by active ballistic
excursions with constant speed and random direction (blue phases moving
along the motor tracks). The duration of both phases is distributed
exponentially with mean times $\tau_{p,a}$. When the particle reaches the
receptor (orange sphere) it binds/dissociates with rates $k_\mathrm{on}$ and
$k_\mathrm{off}$. Due to the specific geometry the system is
effectively 1-dimensional.}
\label{schm}
\end{center}
\end{figure}

In a mean field picture of receptor signalling at equilibrium, developed by Bialek
and Setayeshgar \cite{Bialek}, signalling molecules diffuse in space and reversibly
bind to the receptor in a Markovian fashion (Fig.~\ref{schm}). The central object
of the theory is the so-called receptor-noise correlation time $\tau_c$
\cite{BergPurc,Bialek,Tkacik,GodecActive,Goodhill}. Namely, in a setting where the
receptor measures the concentration over a period $\tau_m$---much longer than any
correlation time in the system---the noise in the receptor occupancy statistic will
be Poissonian, and the concentration estimate will improve with the number $N_i
\propto \tau_m/\tau_c$ of independent measurements. The correlation time is set by
the thermal noise in the binding to the receptor and the thermal diffusion of the
signalling molecules \cite{BergPurc,Bialek,Tkacik,Goodhill} but can be altered by
certain details of the transport, such as intermittent sliding along DNA in the
so-called facilitated diffusion model of gene regulation \cite{Tkacik} and
intermittent active excursion by hitchhiking molecular motors \cite{GodecActive}. 
In addition, $\tau_c$ depends on the dimensionality of the cell or domain in which
it occurs \cite{Tkacik,Goodhill}. Moreover, when molecules explore their
surrounding space in a compact manner---the motion is recurrent in the sense of
returning to already visited sites \cite{Feller}---such as the one observed in
1-dimensional diffusion, the recurrences prolong $\tau_c$ and thus reduce $N_i$
within a given fixed $\tau_m$ \cite{Tkacik}. Conversely, the interaction with a
confining domain disrupts the positional correlations at long times
\cite{GodSciRep} and thereby truncates $\tau_c$, causing an improvement of the
sensing precision especially in low dimensions \cite{Goodhill}. 

It was shown in the case of chemical reactions coupled to active transport
that the effect of active excursions is most pronounced in low dimensions
since they act by disrupting the recurrence of 1-dimensional Brownian motion
\cite{Gleb,RMP,Olivier}. Here we demonstrate that the effect of intermittent
active motion in 1-dimensional diffusive systems is even stronger when it comes
to the sensing precision. We compute analytically the accuracy limit for receptor
mediated concentration measurements in dimension 1 and argue that active excursions
allow for enhanced precision of signalling in neurons.

\section{Linear response theory of receptor noise coupled to active transport}

We consider a signalling molecule (mRNA or protein) diffusing on the real line
and randomly switching between a passive diffusion phase $p$ with diffusivity
$D$ and an active ballistic phase $a$ with velocity $\pm v$, see
Fig.~\ref{schm} and \cite{Olivier,RMP,GodecActive}. The duration of active/passive
phases is exponentially distributed with mean $\tau_{a,p}$. The concentrations of
freely diffusing and motor-bound signalling molecules are $c_p(x,t)$ and $c_a^{
\pm}(x,t)$ and $\pm$ denotes motor-bound signalling molecules moving to the
left/right, respectively. In addition, while passively diffusing the signalling
molecule can reversibly bind to a receptor at $x_0$ in a Markov fashion. In a
mean field description the fractional occupancy $n(t)$ with on/off rates
$k_\mathrm{on/off}$ evolves according to the coupled equations 
\numparts
\label{eqs}
\begin{eqnarray}
\frac{dn(t)}{dt}&=&k_{\mathrm{on}}c_{p}(x_0,t)[1-n(t)]-k_{\mathrm{off}}n(t),
\label{governing}\\
\frac{\partial c_{p}(x,t)}{\partial t}&=&D\partial^2_xc_p+\frac{c_a^+(x,t)
+c_a^-(x,t)}{\tau_a}-\frac{c_p(x,t)}{\tau_p}-\delta(x-x_0)\frac{dn(t)}{dt},
\label{governing2}\\
\frac{\partial c_a^{\pm}(x,t)}{\partial t}&=&\mp v \partial_x c_a^{\pm}(x,t)
-\frac{c_a^{\pm}(x,t)}{\tau_a}+\frac{c_p(x,t)}{2\tau_p},
\label{governing3}
\end{eqnarray}
\endnumparts
where detailed balance is fulfilled for the binding $k_{\mathrm{on}}\langle
c_p\rangle/k_{\mathrm{off}}=\exp(F/k_BT)$ involving the binding free
energy $F$. Eqs.~(\ref{governing})-(\ref{governing3}) describe the
  motion of a molecule randomly switching between phases of passive diffusion and
  ballistic motion with rates $\tau_p^{-1}$ and
  $\tau_a^{-1}$. Once the molecule locates the receptor at $x_0$ while
  being in the passive phase, it can bind to it. The total binding
  rate is proportional to the intrinsic rate
  $k_{\mathrm{on}}$, the probability $c_{p}(x_0,t)$ to find the molecule at $x_0$ in
  the passive phase, and the probability $1-n(t)$ that the receptor is unoccupied. Once being bound to the receptor the molecule unbinds with a
first order rate proportional to the intrinsic unbinding rate
$k_{\mathrm{off}}$ and the probability $n(t)$ to find the receptor
occupied. Note that since $c_{p}$ has units of
  1/length and $n(t)$ is dimensionless, the rates $k_\mathrm{on/off}$
  have different units, i.e. $k_\mathrm{on}$ has the units of
  length/time and $k_\mathrm{off}$ has units of
  1/time.

To obtain a closed equation for the dynamics of $n(t)$ close to
equilibrium, we linearise Eqs.~(\ref{governing})-(\ref{governing3}) around the
respective equilibrium values $\langle n \rangle,\langle c_p \rangle$, and
$\langle c_{a}^{\pm}\rangle$ \cite{Bialek} to obtain, in terms of small
fluctuations, $n(t)=\langle n\rangle+\delta n(t)$ and $c_{p}(x,t)=\langle c_p\rangle
+\delta c_p(x,t)$, and $c_a^{\pm}(x,t)=\langle c_a^{\pm}\rangle+\delta c_a^{\pm}
(x,t)$. Moreover, the detailed balance condition imposes the constraint
$\delta k_{\mathrm{on}}/k_{\mathrm{on}}-\delta k_{\mathrm{off}}/k_{\mathrm{off}}=
\delta F/k_BT$ on the free energy fluctuations. By Fourier transforming in time
and in space, $\hat{\mathcal{F}}_t(t\to\omega)[\cdot]=\int_0^{\infty}\mathrm{e}^{
i\omega t}(\cdot)dt$, $\hat{\mathcal{F}}_x(x\to k)[\cdot]=\int_{-\infty}^{\infty}
\mathrm{e}^{-ikx}(\cdot) dx$, and solving the resulting system of ordinary
equations we arrive at an exact generalised Langevin equation for the fluctuations
around the equilibrium receptor occupancy within the linear regime
\cite{GodecActive}
\begin{equation} 
\label{Langevin}
\int_0^t\gamma(t-t')\frac{d\delta n(t')}{dt'}+\tau_b^{-1}d\delta
n(t)=\frac{k_{\mathrm{off}}\langle n \rangle\delta F(t)}{k_BT}.
\end{equation}
Here $\tau_b$ denotes the correlation time of two-state Markov switching, $\tau_b=
(k_{\mathrm{on}}\langle c_p\rangle-k_{\mathrm{off}})^{-1}$, and the noise in the
form of the free energy fluctuations $\delta F(t)$ has zero mean $\langle\delta
F(t)\rangle=0$ and obeys the fluctuation-dissipation theorem $\langle\delta
F(t)\delta F(t')\rangle=2(k_BT/k_{\mathrm{off}}\langle n \rangle)^2\gamma(t-t')$
\cite{LanLif}. The memory
kernel $\gamma(t)$ in terms of an inverse Fourier transform operator reads
\begin{equation} 
\label{memory}
\gamma(t)=\delta(t) + \hat{\mathcal{F}}^{-1}_t\left[\lim_{a\to0}\int_{-\pi/a}^{
\pi/a}\mathrm{e}^{ik x}\frac{dk}{2\pi}\frac{k_{\mathrm{on}}(1-\langle n\rangle)
}{-i\omega+Dk^2+\Lambda_k(\tau_a,\tau_p;\omega)}\right],
\end{equation}
and the contribution due to the intermittent active excursions is
\begin{equation} 
\label{active}
\Lambda_k(\tau_a,\tau_p;\omega)= (\tau_a\tau_p)^{-1}\frac{\tau_a^{-1}-i\omega}{
v^2k^2+(\tau_a^{-1}-i\omega)^2}.
\end{equation} 
The limit in Eq.~(\ref{memory}) is to be understood as a finite
receptor size taken to zero after the integral is evaluated in order
for the integral to converge. The memory term in the Langevin equation
(\ref{Langevin}) reflects the fact that it takes a finite time before
the receptor feels the effect of $\delta F(t)$ because the signalling molecule
moves throughout space before (re)binding. 

According to linear response theory \cite{Bialek,LanLif} we can write $\delta
n(t)=\int_0^t\alpha(t')\delta F(t-t')dt'$ where the generalised susceptibility
becomes
\begin{equation} 
\label{susc}
\alpha(t)=\hat{\mathcal{F}}^{-1}_t\left[\tilde{\alpha}(\omega)\right]=\hat{
\mathcal{F}}^{-1}_t\left[\frac{\delta\tilde{n}(\omega)}{\delta\tilde{F}(\omega)}
\right],
\end{equation} 
and the power spectrum of $\delta n(t)$ is in turn obtained according to the
fluctuation-dissipation theorem from the imaginary part of $\tilde{\alpha}(
\omega)$,
\begin{equation} 
\label{FDT}
S_{\delta n}(\omega)=\frac{2k_BT}{\omega}\mathrm{Im}[\tilde{\alpha}(\omega)].
\end{equation}    
Since the receptor's sensitivity is limited to frequencies
$|\omega|\le\tau_m^{-1}$, the uncertainty in measuring the occupation fraction
will be 
\begin{equation} 
\label{uncN}
\overline{\delta n^2}=\int_{-1/\tau_m}^{1/\tau_m}S_{\delta n}(\omega)d\omega. 
\end{equation} 
Moreover, a change in concentration is equivalent to a change in $F$,
$\delta c_p/\langle c_p\rangle=\delta F/k_BT$. Using this one can also
show that
\cite{Bialek} 
\begin{equation} 
\label{uncC}
S_{\delta c_p}(\omega)=\left(\frac{\langle  c_p\rangle}{k_BT}\right)^2\left|
\frac{\delta\tilde{n}(\omega)}{\delta\tilde{F}(\omega)}\right|^{-2}S_{\delta
n}(\omega),
\end{equation}
and use this to relate the uncertainty in $\delta n$ to the precision
at which the receptor can determine $c_p$.

\section{Equilibration rate}

We split the signalling process in an equilibration phase, during which the
system equilibrates to a new concentration, and the measurement phase, during
which the receptor reads out this equilibrium concentration. Moreover, we
assume that the equilibration time corresponds to the time during which
the signalling molecules move a distance $L$ of the order of the size of the cell
or a cellular compartment. The equilibration time $\tau_i$ is then defined
implicitly by the mean squared displacement via $\langle x (\tau_i)^2\rangle=L^2$.

\begin{figure}
\begin{center}
\includegraphics[width=16.cm]{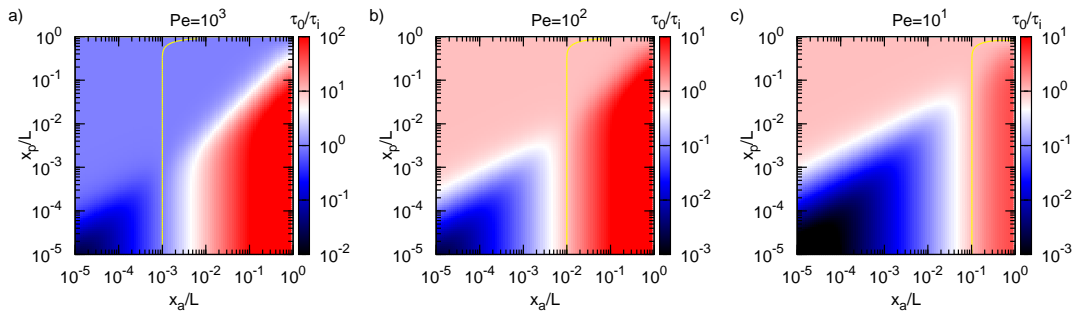}
\caption{Ratio $\tau_0/\tau_i$ of equilibration times for passive
diffusion (subscript $0$) and intermittent active motion (subscript
$i$) as a function of the typical lengths of active ($x_a$) and passive ($x_p$)
displacements for various P{\'e}clet numbers $\mathrm{Pe}=Lv/D$. The
yellow line corresponds to $\tau_0/\tau_i=1$. Whenever
$\tau_0/\tau_i>1$ active motion leads to faster equilibration.}
\label{rate}
\end{center}
\end{figure}

We here neglect the binding to the receptor given by Eq.~(\ref{governing}) and
adopt a probabilistic interpretation of Eqs.~(\ref{governing2})
and (\ref{governing3}), which we solve by Laplace transforming in time and
Fourier transforming in space. The mean squared displacement for a particle
starting at the origin in the passive phase is obtained from the Laplace transform
$\langle x^2(s)\rangle=-\partial_k^2[c_a^{+}(k,s)+c_a^{-}(k,s)+c_p(k,s)]_{k=0}$
and after Laplace inversion reads
\begin{eqnarray}
\langle x(t)^2\rangle&=&2\Bigg\{(v\tau_a)^2e^{-\frac{t}{\tau_a}}-
\frac{v^2+D\tau_p^{-1}}{(\tau_a^{-1}+\tau_p^{-1})^2}e^{-\frac{t(\tau_a+\tau_p)}{
\tau_a\tau_p}}\nonumber \\
&+&\frac{(v\tau_a)^2+D\tau_p}{\tau_p(1+\tau_a/\tau_p)}t+\frac{D\tau_p-(v\tau_a)^2
(1+2\tau_p/\tau_a)}{(1+\tau_p/\tau_a)^2}\Bigg\}.
\label{msd}
\end{eqnarray}
Eq.~(\ref{msd}) is a transcendental equation for $\tau_i$ and depends only on
three parameters: the typical distance covered in the active and passive phases,
$x_a=v\tau_a$ and $x_p=\sqrt{D\tau_p}$, and the dimensionless P{\'e}clet number
$\mathrm{Pe}=Lv/D$. Moreover, it states that over a period of duration $\tau_a
+\tau_p$ the directional persistence in the active phase causes a nonlinear time
dependence of $\langle x^2(t)\rangle$. Upon this transient regime an effective
diffusive regime $\langle x^2(t)\rangle\sim D_{\mathrm{eff}} t$ is established
with an effective diffusion coefficient $D_{\mathrm{eff}}=(D\tau_p+[v\tau_a]^2)/
(\tau_p+\tau_a)$. To estimate the equilibration rate of active transport
with respect to diffusion we compare $\tau_i$ with the purely passive equilibration
time $\tau_0\equiv L^2/(2D)$. Fig.~\ref{rate}a)-c) shows results for various
biologically relevant P{\'e}clet numbers.

From Fig.~\ref{rate} we find that active transport is more efficient for larger
$\mathrm{Pe}$ values. More precisely, the required typical displacement in the
active phase needed to enhance the equilibration with respect to bare diffusion
is smaller for larger $\mathrm{Pe}$. In the biologically relevant setting the
molecular motor speed $v \sim 1\mu\mathrm{m}/\mathrm{sec}$ is widely independent
of the particle size \cite{Alberts} and the values for the diffusion coefficients
span a scale between $D\lesssim 10^{-2}\mu\mathrm{m}^2/\mathrm{sec}$ corresponding
to large cargo such as vesicles, and $D\sim10\mu\mathrm{m}^2/\mathrm{sec}$
corresponding to smaller proteins. Conversely, the dimension of effectively linear
cells such as neurons or their sub-structures (i.e. dendrites) falls between $10
\mu\mathrm{m}$ and $\lesssim1\mathrm{m}$, which means that $\mathrm{Pe}\gtrsim
10-100$ values are in fact robustly expected. Therefore, according to
Fig.~\ref{rate} it is quite plausible that intermittent active motion indeed
enhances signalling speed in vivo.  

The physical principle underlying the enhancement is rooted in the fundamental
difference in the time scaling of diffusive and active motion, $\simeq t$ versus
$\simeq t^2$. For example, comparing only purely passive and active motion we
find that for $\mathrm{Pe}>2$ active motion is more efficient. In the intermittent
case the motion has a transient period of duration $\tau_a+\tau_p$, which
corresponds to a parameter dependent combination of both regimes. After this
transient period the effective diffusive regime is established with diffusivity
$D_{\mathrm{eff}}$, which may or may not be larger than the bare $D$. $\tau_i$
can therefore be smaller or larger than $\tau_0$. Shuttling of large cargo
therefore almost universally profits from active motion, whereas active motion
of smaller proteins will only be more efficient over sufficiently large distances.
The observed features thus provide a simple explanation why experimentally active
transport is observed mostly in the trafficking of larger particles
\cite{BialekBK,Kinesins}. Similarly, active diagnostics
\cite{Diagnostics,Diagnostics2} can also
be faster and hence could enable for a higher diagnostic throughput.

\section{Signalling precision with thermal diffusion alone}
 
We now address the signalling precision and focus first on the
situation, where molecules move in space by thermal diffusion alone.
In this case $\Lambda_k=0$ and the $k$-integral in Eq.~(\ref{memory})
is evaluated exactly, after taking the limit $a\to 0$ yielding
\begin{equation} 
\label{mem_P}
\gamma(t)=\delta(t)+\hat{\mathcal{F}}^{-1}_t\left[\frac{k_{\mathrm{on}}(1-\langle n\rangle)}{2}\sqrt{\frac{-i}{D\omega}}\right].
\end{equation}
Using Eq.~(\ref{mem_P}) in 
Eqs.~(\ref{FDT}) to (\ref{uncC}) we arrive at the power spectrum of
concentration fluctuations experienced by the receptor,
\begin{equation} 
\label{unc_p}
S_{\delta c_p}(\omega)=\frac{2\langle c_p\rangle}{k_{\mathrm{on}}(1-\langle n
\rangle)}-\frac{\langle c_p\rangle}{\sqrt{D}}\frac{\sqrt{|\omega|}}{\omega}\sin
\left(\frac{\mathrm{Arg}(-i\omega)}{2}\right),
\end{equation}
where $\mathrm{Arg}$ denotes the principal value of the argument. Integrating
over the frequency range $(-\tau_m^{-1},\tau_m^{-1})$ we obtain the final result
for the variance of the concentration measured by the receptor,
\begin{equation} 
\label{uncC_p}
\overline{\delta c_p^2}=\frac{2\langle c_p\rangle}{k_{\mathrm{on}}(1-\langle n\rangle)\tau_m}+\frac{\langle c_p\rangle}{\pi}\sqrt{\frac{2}{D\tau_m}},
\end{equation}
where the first part describes the noise due to the two-state Markov switching
(i.e. the binding alone) and the second term stands for the noise due to diffusion.
Note that for the recurrent nature of 1-dimensional diffusion and the fact that the
receptor is point-like, we \emph{cannot} approximate the precision at which the
receptor can determine $c_p$ with $\int_{-1/\tau_m}^{1/\tau_m}S_{\delta c_p}(
\omega)d\omega\sim S_{\delta c_p}(\omega\to 0)/\tau_m$ as in the 3-dimensional
case (see e.g. \cite{Bialek}). More precisely, in contrast to the Lorentzian shape
of $S_{\delta c_p}(\omega\to 0)$ in the 3-dimensional case, $S_{\delta c_p}(\omega)
$ diverges as $\omega\to 0$. The integral over $\omega$ nevertheless converges
and leads to Eq.~(\ref{unc_p}). Moreover, in contrast to the 3-dimensional case
where the squared measurement error $\overline{\delta c_p^2}$ decreases as $1/
\tau_m$, for 1-dimensional diffusion we find the much slower decay $\overline{\delta
c_p^2}\propto1/\sqrt{\tau_m}$. That is, $N_i^{1d}/N_i^{3d}\propto 1/\sqrt{\tau_m}$
and the receptor measurement is thus much less efficient in 1-dimension.

\section{Signalling precision with active motion}

As we are interested in the signalling precision at equilibrium and hence consider
$\tau_m$ values which are much longer than any correlation time in the motion
\cite{Bialek,Tkacik,GodecActive} such that $\tau_m\gg\tau_a,\tau_p$, we may take
the limit in $\Lambda_k(\tau_a,\tau_p;\omega \to 0)$ as well as in
Eq.~(\ref{memory}). This way we recover, after performing the integral
over $k$ in Eq.~(\ref{memory}) and taking the limit $a\to 0$,
an effective white noise asymptotic on the slow time scale $t\gg \tau_a,\tau_p$,
\begin{equation} 
\label{mem_A}
\gamma(t)\sim \delta(t)\left[1+\frac{k_{\mathrm{on}}(1-\langle n\rangle)}{D^2\tau_p\left([D\tau_p]^{-1}+[v\tau_a]^{-2}\right)^{3/2}}\right],
\end{equation} 
and correspondingly an effectively Lorentzian
fluctuation spectrum $S_{\delta n}(\omega)$ at small frequencies (see
\cite{GodecActive}). From Eq.~(\ref{uncC}) we obtain also the low frequency
region of the power spectrum concentration fluctuations,
\begin{equation} 
\label{uncA}
S_{\delta c_p}(\omega)\sim\frac{2\langle c_p\rangle}{k_{\mathrm{on}}(1-\langle
n\rangle)}+\frac{\langle c_p\rangle}{D x_p^2\left(x_p^{-2}+x_a^{-2}\right)^{3/2}},
\end{equation}
for $\omega\ll\tau_b^{-1},\tau_a^{-1},\tau_p^{-1}$, where we introduced the typical
distance the signalling molecule moves in the passive $x_p=\sqrt{D\tau_p}$ and
motor bound phases $x_a=v\tau_a$. As before, the first term in Eq.~(\ref{uncA})
corresponds to the two-state switching noise and the second term to the noise due
to spatially extended intermittent dynamics. Note that in contrast to the
3-dimensional setting, where the active excursions merely rescale the correlation
time \cite{GodecActive}, we here find a qualitative change in the properties of
the noise, compare Eqs.~(\ref{unc_p}) and (\ref{uncA}). 

Using Eq.~(\ref{uncA}) we can now approximate the precision at which the receptor
can determine $c_p$ with $\int_{-1/\tau_m}^{1/\tau_m}S_{\delta c_p}(\omega)d\omega
\sim S_{\delta c_p}(\omega\to 0)/\tau_m$ and obtain our main result
\begin{equation} 
\label{uncC_a}
\overline{\delta c_p^2}\sim\frac{2\langle c_p\rangle}{k_{\mathrm{on}}(1-\langle
n\rangle)\tau_m}+\frac{\langle c_p\rangle}{D x_p^2\left(x_p^{-2}+x_a^{-2}\right)
^{3/2}\tau_m}.
\end{equation}
Here we are interested in the transport-controlled sensing
\cite{Bialek,Tkacik,GodecActive}. Comparing the noise due to the spatially
extended motion for passive and active intermittent motion we find that that
active motion allows for more precise absolute concentration measurements as
soon as the inequality
\begin{equation} 
\label{ineq}
\tau_m > \tau_p \frac{\pi^2}{2(1+(x_p/x_a)^{2})^{3}}
\end{equation}
holds such that in the limit of long active excursions $x_a\gg x_p$ we end up
with the condition $\tau_m>\tau_p \frac{\pi^2}{2}$. Note that the right hand
side of this inequality is essentially the characteristic time of the asymptotic
exponential decay of the first passage time density of a 1-dimensional random
walk in a domain of length $L$ if we set $L^2/D=\tau_p$ \cite{Redner}. In other
words, for active signalling to be more precise in 1-dimension the receptor needs
to measure long enough for the particle to find the target in the passive phase,
which is an intuitive result.

\begin{figure}
\begin{center}
\includegraphics[width=16.cm]{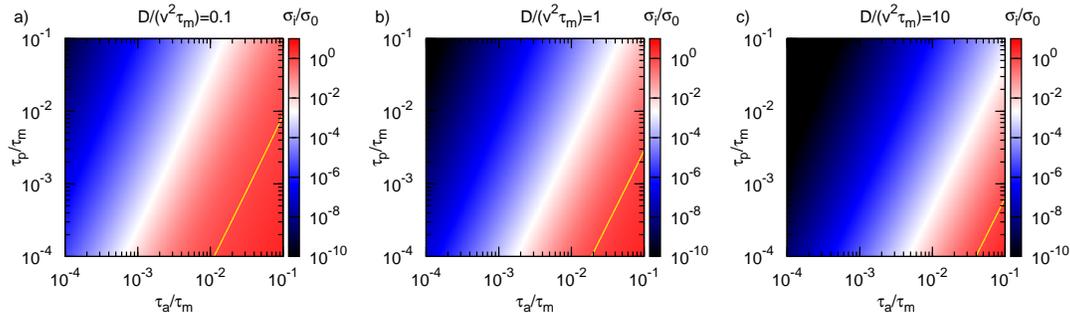}
\caption{Precision ratio of scaled variances of $\sigma_k=\overline{\delta c_{p,k}
^2}/\langle c_{p,k}\rangle^2$ with $k=0,i$ for active intermittent (subscript $i$)
versus passive (subscript $0$) transport as a function of the relative duration of
active ($\tau_a/\tau_m$) and passive ($\tau_a/\tau_m$) phases with respect to the
measurement time $\tau_m$ for various values of dimensionless ratio between the
squared typical length of passive $x^2_{p,m}=D\tau_m$ versus active
$x^2_{a,m}=(v\tau_m)^2$ displacements during the measurement time
$\tau_m$. Whenever $\sigma_i/\sigma_0<1$ active motion leads to more
precise signalling. Note that $\tau_a/\tau_m\le 0.1$ and
$\tau_p/\tau_m\le 0.1$ in order to assure equilibrium sensing conditions.}
\label{precision}
\end{center}
\end{figure}

In order to be more concrete we compare the scaled variances of measurement
errors for active $\sigma_i=\overline{\delta c_{p,i}^2}/\langle c_{p,i}\rangle^2$
and passive $\sigma_0=\overline{\delta c_{p,0}^2}/\langle c_{p,0}\rangle^2$
motion. In the transport-controlled regime we have $\langle
c_{p,i}\rangle\sim c_{\mathrm{tot}}/(1+\tau_a/\tau_p)$ for
intermittent active motion and $\langle c_{p,0}\rangle \sim
c_{\mathrm{tot}}$, where $c_{\mathrm{tot}}$ denotes the total
concentration of molecules. Note that here and throughout the entire paper we
implicitly assume that the number of molecules exceeds the number of
receptors \cite{Bialek}. The relative precision ratio reads
\begin{equation}
\frac{\sigma_i}{\sigma_0}=\pi\sqrt{\frac{\tau^{\ast}_p}{2}}\frac{(1+\tau^{\ast}_a/
\tau^{\ast}_p)}{\left(1+Q\tau^{\ast}_p/[\tau^{\ast}_a]^2\right)^{3/2}},
\label{prec_rat}
\end{equation}
where we introduced dimensionless times $\tau^{\ast}_p=\tau_p/\tau_m$
and $\tau^{\ast}_a=\tau_a/\tau_m$ as well as $Q=D(v^2\tau_m)$, the
dimensionless ratio between the squared typical lengths of passive
$x_{p,m}=\sqrt{D\tau_m}$ versus active
$x_{a,m}=v\tau_m$ displacements during the measurement time
$\tau_m$. The results for various values of $Q$ are presented in
Fig.~\ref{precision}. 

We find that the minimal value of $\tau_a$ that is required for improved
sensing precision with respect to bare diffusion (i.e. for
$\sigma_i/\sigma_0<1$, which corresponds to the region to the left of
the yellow curve in Fig.~\ref{precision}) decreases with decreasing
$Q$. In other words, for large particles with a smaller $D$ the
active displacements can become arbitrarily short. Given that the
typical measurement times lie between $\simeq 1$ sec and $\simeq 1$ min
\cite{Bialek} the conditions for improved signalling accuracy appear
to be robustly satisfied. 

To understand this
we need to recall that, while larger $\tau_a$ monotonically leads to lower
absolute read-out errors (see Eq.~(\ref{uncC_a})), it simultaneously
decreases $\langle c_p\rangle$ and hence renormalises $\sigma_i$. The
improved accuracy in Fig.~\ref{precision} is thus a result of a
trade-off between a decreases of the absolute concentration fluctuations and a
lower equilibrium probability to be at the receptor site. This result
is striking as it suggests that even the slightest active
displacements can disrupt the recurrence and improve the
read-out precision as long as their length is larger than the receptor
size.

Physically, this observation is due to the fact that the receptor
collects new information only from statistically independent binding
events. Correlations between consecutive measurements arise due to
a finite Markov binding time $\tau_b$ and due to the
return and rebinding of a previously bound molecule. Moreover, we
assume that only
freely diffusing molecules can bind to the receptor. Therefore, the
receptor necessarily experiences the binding of those
molecules, which are ballistically swept towards the binding site
over a distance larger than the receptor size, as statistically
independent. In turn, molecules which are ballistically flushed away
from the receptor after unbinding will also contribute statistically
independent binding events, regardless of how they return to the receptor. 
The non-existence of a lower-bound on $\tau_a$ is thus an
artefact of assuming a point-like receptor.

Note that in an alternative setting, in which we compare the precision to
determe the same concentration of passively moving molecules, which corresponds
to a higher $c_{\mathrm{tot}}$ in the intermittent active case
(i.e. $\langle c_{p,i}\rangle\to c_{\mathrm{tot}}$ \cite{GodecActive}),
the signalling precision would be improved unconditionally.   
Therefore, in contrast to the 3-dimensional case, where active motion only
improves sensing precision for certain values of parameters
\cite{GodecActive}, active transport can robustly and much more
efficiently improve sensing accuracy in 1-dimensional systems for
sufficiently long measurement times.

\section{Conclusion}

The degree of recurrence of spatial exploration is essential for random target
search processes \cite{Gleb,RMP}. For example, in the facilitated diffusion model
of gene regulation the topological coupling of 1- and
3-dimensional diffusion allows for a more efficient search
(e.g. \cite{gene}). In a similar manner intermittent active
excursions can significantly speed up random search \cite{Gleb,RMP}.    

In contrast, the topological coupling of 1- and 3-dimensional diffusion does not
appreciably improve the signalling precision \cite{Tkacik}. In addition, we showed
previously that in a 3-dimensional setting active motion only conditionally
improves the signalling accuracy, by decreasing the correlation time of the
counting noise in a process called active focusing
\cite{GodecActive}. Here we find, strikingly, that active excursions
effect qualitative changes in the power spectrum of concentration
fluctuations experienced by the receptor in
1-dimensional systems such as neurons. By adding the active component
the power spectrum changes from $1/\sqrt{\omega}$ for thermal
diffusion alone to a Lorentzian shape with a finite plateau. This Lorentzian
shape is also observed for passive signalling in 3-dimensions
\cite{Bialek,Tkacik,GodecActive}. Therefore, active excursions
disrupt the recurrent nature of 1-dimensional diffusion.

Existing studies provide insight into how receptor clustering \cite{Bialek} and
cooperativity \cite{Bialek2}, dimensionality \cite{Tkacik}, spatial confinement
\cite{Goodhill}, receptor diffusion \cite{Goodhill2} and active transport
\cite{GodecActive} affect the precision of receptor signalling. The overall
dependence of the counting noise on the manner the signalling molecules explore
their surrounding space suggests that a heterogeneous diffusivity profile
\cite{GodSciRep,heterogen} and spatial disorder \cite{disorder} would alter
the signalling precision as well. Both have been observed in experiments
\cite{hetexp}. In addition, signalling molecules or transport versicles often
exhibit anomalous diffusion \cite{anomalous}, both in the form of passive
\cite{passive} and active \cite{christine,active} motion. It would therefore be
interesting to investigate the impact of these features on the sensing precision
in the future.

\ack

AG acknowledges funding through an Alexander von Humboldt Fellowship
and ARRS project Z1-7296.

\begin{appendix}

\end{appendix}


\section*{References}

\begin{thebibliography}{99}

\bibitem{Smol} von Smoluchowski M 1916 \emph{Phys. Z.} \textbf{17}, 557.

\bibitem{BergPurc} Berg H C and Purcell E M 1977 \emph{Biophys. J.} \textbf{20} 193.

\bibitem{Bialek} Bialek W and Setayeshgar S 2005 \emph{Proc. Natl. Acad. Sci. USA}
\textbf{102} 10040.

\bibitem{Tkacik} Tka\v cik G and Bialek W 2009 \emph{Phys. Rev. E} \textbf{79}, 051901.

\bibitem{Endres} Endres R G and Wingreen N S 2008 \emph{Proc. Natl. Acad. Sci. USA}
\textbf{105}, 15749;\\ \noindent Rappel W-J and Levine H 2008 \emph{Phys. Rev. Lett.} \textbf{100}, 228101;\\ \noindent Hu B, Kessler D A, Rappel W-J, and  Levine H 2011 \emph{Phys. Rev.
Lett.} \textbf{107}, 148101;\\ \noindent Govern C and ten Wolde P R 2012 \emph{Phys. Rev. Lett.} \textbf{109},
218103;\\ \noindent Kaizu C et al. 2014 \emph{Biophys. J.} \textbf{106}, 976;\\ \noindent
Tka\v cik G, Gregor T, and Bialek W 2008 \emph{PLoS ONE} \textbf{3},
e2774.

\bibitem{GodecActive} Godec A and Metzler R 2015 \emph{Phys. Rev. E} \textbf{92} 010701(R).

\bibitem{Goodhill} Bicknell B A, Dayan P, and  Goodhill G J 2015, \emph{Nat. Commun.} \textbf{6} 7468.

\bibitem{Diagnostics} Korten T, M\aa nsson A, and Diez S 2010 \emph{Curr. Opin.
Biotechnol.} \textbf{21}, 477.
 
\bibitem{Diagnostics2}  Hess H and Vogel V 2001 \emph{Rev. Mol. Biotechnol.}
\textbf{82}, 67. 

\bibitem{Stoch} Li G-W and Xie X S 2011 \emph{Nature} \textbf{475}, 308. 

\bibitem{Xie} Gebhardt J C M et al. 2013 \emph{Nat. Methods} \textbf{10}, 421.

\bibitem{Elf1} Persson F, Lind\'en M, Unoson C, and Elf J 2013 \emph{Nat. Methods}
\textbf{10}, 265.

\bibitem{Elf2} Hammar P et al. 2014  \emph{Nat. Genetics} \textbf{46} 405.

\bibitem{BialekBK} Bialek W, \emph{Biophysics: Searching for Principles}
(Princeton University Press, New Jersey, 2012).

\bibitem{Alberts} Alberts B et al., \emph{Molecular Biology of the Cell}
(Garland, New York, 2002).

\bibitem{Motors1} Kolomeisky A B and Fischer M E 2007 \emph{Annu. Rev. Phys. Chem.}
\textbf{58}, 675.

\bibitem{Motors2} J\"ulicher F, Ajdari A, and Prost J 1997  \emph{Rev. Mod. Phys.}
\textbf{69}, 1269.

\bibitem{Interm1} Salman H et al., \emph{ Biophys. J.} \textbf{89}, 2134
  (2005);\\ \noindent Huet S, Karatekin E, Tran V S, Cribier S and Henry J
  P 2006 \emph{Biophys. J.} \textbf{91}, 3542;\\ \noindent Vermehren-Schmaedick A et
  al. 2014 \emph{PLoS ONE} \textbf{9}, e95113;\\ \noindent Arcizet D, Meier B, Sackmann E, R{\"a}dler J O, and
  Heinrich D 2008  \emph{Phys. Rev. Lett.} \textbf{101}, 248103.

\bibitem{mrna} St Johnston D 2005 \emph{Nature Rev.} \textbf{6}, 363;\\ \noindent Tekotte H
and Davis I 2002 Trends in Genet. \textbf{118}, 636; Fusco D et al 2003 \emph{Curr.
Biol.} \textbf{13}, 161;\\ \noindent Vale R D 2003 \emph{Cell} \textbf{112}, 467.

\bibitem{haim} Mussel M, Zeevy K, Diamant H, and Nevo U 2014 \emph{Biophys. J.}
\textbf{106}, 2710;\\ \noindent Roy S et al 2007 \emph{J. Neurosci.} \textbf{27}, 3131;
Scott  D A et al. 2011 \emph{ Neuron.} \textbf{70}, 441. 

\bibitem{streaming} Goldstein R E, van de Meent J-W 2015 \emph{Interface Focus}
\textbf{5}, 20150030.

\bibitem{christine} Reverey J F, Jeon J-H, Bao H, Leippe M, Metzler R and
Selhuber-Unkel C. 2015  \emph{Sci. Rep.} \textbf{5}, 11690. 

\bibitem{Gleb} Oshanin G, Lindenberg K, Wio H~S, and Burlatsky S
2009 \emph{J. Phys. A: Math. Theor.} \textbf{42}, 434008. 

\bibitem{RMP} B{\'e}nichou O, Loverdo C, Moreau M, and Voituriez R 2011.
\emph{Rev. Mod. Phys.} \textbf{83}, 81.

\bibitem{Feller} Feller W 1986 \emph{An Introduction to Probability
  Theory and Its Applications vol. II.} (Wiley, New York).

\bibitem{GodSciRep} Godec A and Metzler R 2016 \emph{Sci. Rep.} \textbf{6},
  20349.

\bibitem{Olivier} B{\'e}nichou O, Loverdo C, Moreau M, and Voituriez R 2008
\emph{Nat. Phys.} \textbf{9} 134;\\ \noindent 2009 \emph{J. Stat. Mech.} P02045.

\bibitem{LanLif} Landau L D and Lifshitz E M 1980 \emph{Statistical
  Physics: Part I} (Pergamon Press, Oxford).  

\bibitem{Kinesins} Hirokawa N, Noda Y, Tanaka Y, and Niwa S 2009 \emph{Nat. Rev. Mol.
Cell Biol.} \textbf{10}, 682;\\ \noindent Desnos C and Huet S 2007 \emph{Biol. Cell.} \textbf{99},
411.

%\bibitem{Olivier} B\'enichou O and Voituriez R 2014 \emph{Phys. Rep.} \textbf{539}, 225.


\bibitem{Redner} Redner S, \emph{A guide to first passage processes.}
Cambridge University Press, New York, 2001.


\bibitem{gene} Hippel PH and Berg OG 1989 J. Biol. Chem. \textbf{264}, 675;\\
Sheinman O, B\'enichou O, Kafri Y, and Voituriez R
2012 \emph{Rep. Prog. Phys.} \textbf{75}, 026601;\\
Pulkkinen O and Metzler R 2013 \emph{Phys. Rev. Lett.} \textbf{110}, 198101;\\
Bauer M and Metzler R 2012 \emph{Biophys. J.} \textbf{102}, 2321;\\
Bauer M and Metzler R 2013 \emph{PLoS ONE} \textbf{8}, e53956;\\
Koslover E F, D\'iaz de la
Rosa M A D, and Spakowitz A J 2011 \emph{Biophys. J.} \textbf{101}, 856;\\
Kolomeisky A 2011 \emph{Phys. Chem. Chem. Phys.} \textbf{13}, 2088;\\
Wunderlich Z and Mirny L A 2008 \emph{Nucleic Acids Res.} \textbf{36},
3570.

\bibitem{Bialek2} Bialek W and Setayeshgar S 2008 \emph{Phys. Rev. Lett.} \textbf{100},
258101.

\bibitem{Goodhill2} Nguyen H, Dayan P, and Goodhill G J 2014 \emph{J. R. Soc. Interface} \textbf{12}, 20141097.

\bibitem{heterogen} Godec A and Metzler R 2015 \emph{Phys. Rev. E}
  \textbf{91}, 052134;\\ \noindent Viccario G, Antoine C, and Talbot J 2015 \emph{Phys. Rev. Lett.}
  \textbf{115}, 240601;\\ \noindent Cherstvy A G, Chechkin A V, and Metzler R
  2014 \emph{J. Phys. A: Math. Theor.} \textbf{47},  485002.

\bibitem{disorder} Sabhapandit S, Majumdar S N, and Comtet A 2006 \emph{Phys. Rev. E}
  \textbf{73} 051102;\\ \noindent Majumdar S N, and Comtet A 2002 \emph{Phys. Rev. Lett.}
  \textbf{89} 060601;\\ \noindent Burov S and Barkai E 2007 \emph{Phys. Rev. Lett.}
  \textbf{98} 250601;\\ \noindent Dean D S, Gupta S, Oshanin G, Rosso
  A, and Schehr G 2014 \emph{J. Phys. A: Math. Theor.}  \textbf{47}, 372001;
  \\ \noindent Kr\"usemann H, Godec A, and Metzler R. 2014 \emph{Phys. Rev. E}
  \textbf{89}, 040101(R);\\ \noindent Kr\"usemann H, Godec A, and Metzler
  R. 2015 \emph{J. Phys. A: Math. Theor.}  \textbf{48}, 285001;\\ \noindent Godec A, Chechkin A V, Barkai E, Kantz H and
  Metzler R 2014 \emph{J. Phys. A: Math. Theor.} \textbf{47}, 492002

\bibitem{hetexp} English B P, Hauryliuk V, Sanamrad A, Tankov S,
  Dekker N H, and Elf J 2011 \emph{Proc. Natl. Acad. Sci. USA} \textbf{108}, E365;\\ \noindent
Cutler P J, Malik M D, Liu S, Byars J S, Lidke
  D S, and Lidke K A 2013 \emph{PLoS ONE} \textbf{8}, e64320
(2013).

\bibitem{anomalous} Barkai E, Garini Y, and Metzler R 2012 \emph{Phys. Today}
  \textbf{65}, 29;\\ \noindent
Metzler R, Jeon J-H, Cherstvy A G, and Barkai E 2014 \emph{Phys. Chem. Chem.
Phys.} \textbf{16}, 24128.

\bibitem{passive} Di Rienzo C, Piazza V, Gratton E, Beltram F, and
Cardarelli F 2014 \emph{Nature Commun.} \textbf{5}, 5891.;\\
Jeon J-H, Tejedor V, Burov S, Barkai E, Selhuber-Unkel C, Berg-S{\o}rensen K,
Oddershede L and Metzler R 2011 \emph{Phys. Rev. Lett.} \textbf{106}, 048103;\\
Golding I and Cox E C 2006 \emph{Phys. Rev. Lett.} 96, 098102.

\bibitem{active} Caspi A, Granek R, and Elbaum M 2002 \emph{Phys. Rev. E} \textbf{66},
011916;\\
Gal N and Weihs D 2010 \emph{Phys. Rev. E} \textbf{81}, 020903(R);\\
Goychuk I, Kharchenko V O, and Metzler R 2014 \emph{Phys. Chem. Chem. Phys.}
\textbf{16}, 16524 (2014);\\
Seisenberger G, Ried MU, Endre{\ss} T, B{\"u}ning H, Hallek M and Br{\"a}uchle
C 2001 \emph{Science} \textbf{294}, 1929.

\end{thebibliography}
\end{document}